\documentclass[aps,pre,reprint,groupedaddress]{revtex4-1}

\usepackage{graphicx}
\usepackage{amssymb,amsfonts,amsmath}

\begin{document}

\title{Optimal Dimensionality Reduction of Complex Dynamics:\\ The Chess Game as Diffusion on a Free Energy Landscape.}

\author{Sergei V. Krivov}

\affiliation{Institute of Molecular and
    Cellular Biology, University of Leeds, UK}

\date{\today}

\begin{abstract}
Dimensionality reduction is ubiquitous in analysis of complex dynamics. The conventional dimensionality reduction techniques, however, focus on reproducing the underlying configuration space, rather than the dynamics itself. The constructed low-dimensional space does not provide complete and accurate description of the dynamics. Here I describe how to perform dimensionality reduction while preserving the essential properties of the dynamics. The approach is illustrated by analyzing the chess game - the archetype of complex dynamics. A variable that provides complete and accurate description of chess dynamics is constructed. Winning probability is predicted by describing the game as a random walk on the free energy landscape associated with the variable. The approach suggests a possible way of obtaining a simple yet accurate description of many important complex phenomena. The analysis of the chess game shows that the approach can quantitatively describe the dynamics of processes where human decision-making plays a central role, e.g., financial and social dynamics.
\end{abstract}

\maketitle

\section*{Introduction.}
Complex processes are often described by a single variable to simplify their analyses. Examples include order parameters in physics \cite{hedges_dynamic_2009, krivov_hidden_2004}, biomarkers in medicine \cite{sawyers_cancer_2008,singh_biomarkers_2009,poste_bring_2011}, indexes and asset prices in economics \cite{victor_questioning_2010,haldane_systemic_2011}, citation counts in bibliometrics \cite{hirsch_index_2005,fersht_most_2009}, and educational grades.
To have descriptive and predictive power, or serve as an optimization target, the variable should completely specify the current state and future dynamics of the process. Construction of such variables is challenging. Conventional dimensionality reduction techniques, such as principal component analysis, and their generalizations  \cite{tenenbaum_global_2000,roweis_nonlinear_2000} often fail to produce such variables \cite{krivov_hidden_2004}. The techniques focus on compact representation of an ensemble of configurations (the configuration space). The dynamical information contained in the temporal sequence of the configurations (trajectory) is ignored.

If the future dynamics is completely specified by the current value of the variable and does not depend on history or the values of other variables, the dynamics is said to be Markovian. In this case the dynamics of the variable reproduces the original dynamics, i.e., the projection preserves the dynamics. The stochastic dynamics of the variable can be simply described as diffusion on a free energy profile and is completely specified by the free energy profile and the diffusion coefficient. Conversely, diffusive dynamics implies Markovianity.

Here I describe a method of constructing variables (the optimal reaction coordinate hereafter) which preserve the dynamics upon the projection. The method originated in the protein folding field, where the reaction coordinate and the associated free energy landscapes are used to described in a simplified yet accurate way 
the complex dynamics of protein folding (see inset of Fig. 1) \cite{onuchic_protein_1996,dobson_protein_1998}. The optimal reaction coordinate is constructed based on the system dynamics. Given reaction coordinate time series $X(i\Delta t)$, the cut-based free energy profile (cFEP) can be constructed
$F_C(x)/kT=-ln Z_C(x)$, where partition function  $Z_C(x)$ equals half the number of transitions performed by the reaction coordinate times series through point x \cite{krivov_diffusive_2008, krivov_is_2010} (for details see Appendix). The cFEP is complementary and superior to the conventional histogram based free energy profile as it is invariant to reaction coordinate rescaling, insensitive to statistical noise and capable of detecting sub-diffusion. Together they determine the coordinate dependent diffusion coefficient $D(x)$ and thus completely specify diffusive dynamics \cite{krivov_diffusive_2008}. The optimal reaction coordinate is the one with the highest cFEP \cite{krivov_diffusive_2008, krivov_is_2010} with the following rationale. It generalizes the definition of the transition state as the minimum cut to any position on the reaction coordinate \cite{krivov_one-dimensional_2006}. Projection on a ''bad'' reaction coordinate results in smaller barriers and faster kinetics due to overlapping of different parts of the configuration space. The optimal coordinate exhibits the slowest kinetics \cite{krivov_diffusive_2008}. The dynamics projected on this coordinate is closest to diffusive \cite{krivov_is_2010}. A putative functional form of the reaction coordinate is proposed based on a general understanding of the process. For example, a linear combination of "features" that could describe the process. For protein folding it could be a weighted sum of distances between atoms of a protein \cite{krivov_is_2010}. The coordinate is optimized (trained) on a sample of trajectories representing realizations of the process. The coefficients of the functional form are numerically optimized to make the cFEP  along the coordinate the highest. 

\begin{figure}[htbp]
\centering 
\resizebox*{\columnwidth}{!}{\includegraphics*[]{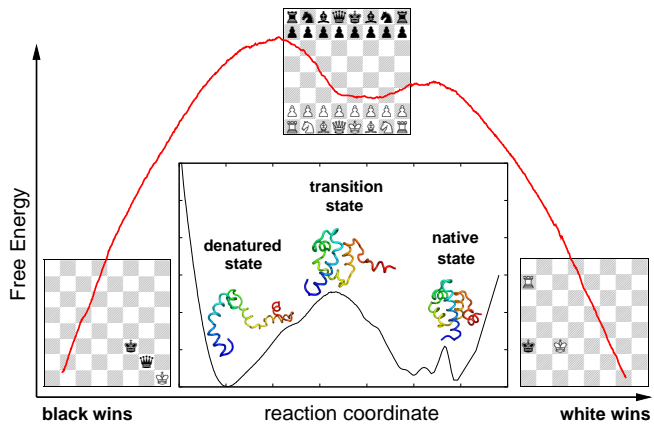}}

\caption{ (color online) \textbf{Cartoon chess free energy landscape.} The game is described as a random walk (diffusion) on the free energy landscape (red line). Starting from the middle the game continues until either the right (white wins) or the left (black wins) end of the profile has been reached. Lower barrier for white reflects that white has more chances to win. The boards show representative positions for the regions on the landscape. The inset shows protein folding free energy landscape.}
\label{cartoon}
\end{figure}

To emphasize the power and generality of the approach the dynamics of the chess game is analyzed \cite{blasius_zipfs_2009}. The chess is a model system of research in artificial intelligence. Its complex dynamics is not generated by a physical system and thus applicability of the free energy landscape framework is not evident. The games played by the computer program GNUCHESS (http://www.gnu.org/software/chess) against itself are analyzed here. No generality is lost, since computers surpassed humans at chess when the Deep Blue won the rematch with the World Chess Champion Garry Kasparov (http://www.research.ibm.com/deepblue). The idea that the chess game can be described by a random walk may seem rather extravagant (Fig. 1). To gain an advantage, players devise precise sequences of moves, which are anything but random. However, in contrast to checker \cite{schaeffer_checkers_2007} the chess are not solved yet. One can not tell the result of a chess game starting from any position if played optimally. The result can only be guessed, suggesting applicability of stochastic description to the chess game, which is corroborated by results presented below.

To keep the analysis one-dimensional only games that end in a victory are considered. 10000 games, initially preprocessed (see Appendix) are used for the analysis.
As a putative reaction coordinate the evaluation function $E(p)$, used in computer chess, is chosen. The function gives a quantitative estimation of the value of a position (p) and is a weighted sum of various factors (the major being the material factor) \cite{shannon_programmingcomputer_1950}. For example, a pawn has a material value of 100 and a queen of 1100, so that $E(p)=100(p_w-p_b)+1100(Q_w-Q_b)+ ...$, where $p_w$, $p_b$ and $Q_w$, $Q_b$ are the number of white and black pawns and queens on the board, respectively, and ... includes other factors describing more subtle properties of a position, such as, board control, mobility, pawn structure, passed pawns, etc. Note that alternative functional forms of reaction coordinate (e.g., the artificial neural networks) can be employed.

\begin{figure*}[htbp]
\centering 
\resizebox*{2.0 \columnwidth}{!}{\includegraphics*[]{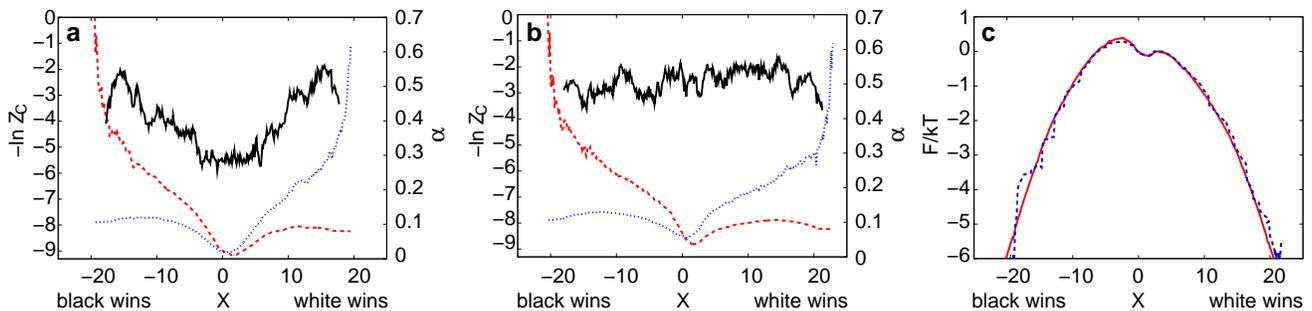}}

\caption{ (color online) \textbf{Free energy profiles of the chess game.} \textbf{a)}  $Z_C^+$ (dashed red), $Z_C^-$ (dotted blue) and $\alpha$ (solid black) for the sub-optimal $E(p)$ reaction coordinate. \textbf{b)} $Z_C^+$ (dashed red), $Z_C^-$ (dotted blue) and $\alpha$ (solid black) for the optimal $\tilde{E}(p)$ reaction coordinate. \textbf{c)} The chess game is described as diffusion on the (equilibrium) cFEP. The cFEP is computed with the free energy profile (solid red) and the Markov network (dashed blue) frameworks (from the cFEPs on panel b). $E(p)$ and $\tilde{E}(p)$ are transformed (rescaled) to X so that the diffusion coefficient $D(x)$ equals to unity.}
\label{joint}
\end{figure*}

\section*{Reaction coordinate optimization.}
The chess game has a non-equilibrium dynamics. The games proceed from the starting position to a checkmate and never backwards. In this case the dynamics is completely specified by $Z_C^+(x)$ and $Z_C^-(x)$ which measure the flux (number of transitions through point $x$) in the positive and negative directions, respectively (Fig. 2a). Their difference manifests the non-equilibrium character of the dynamics and equals (for positive $x$) to the number of games won by white. The exponent $\alpha$ shows that dynamics along $E(p)$ is sub-diffusive $\alpha \sim 0.3$. The exponent measures how the amplitude of the random jumps (changes of the position) scales with time  $\Delta x \sim \Delta t^{\alpha}$; for diffusive dynamics $\alpha=0.5$ and $\Delta x \sim \sqrt{\Delta t}$. The sub-diffusive dynamics indicates that the projected dynamics is not Markovian (consequent displacements are anti-correlated) and that the putative reaction coordinate is not optimal, i.e., its value alone does not completely specify the dynamics. 

It has been shown that sub-diffusive dynamics in protein folding is observed when a sub-optimal reaction coordinate is used for description \cite{krivov_is_2010}. Indeed, the evaluation function $E(p)$ employed in computer chess is a poor reaction coordinate. It can distinguish between positions where white (or black) has a clear advantage. Positions with more subtle advantage or at highly dynamic phases of the game (e.g., during an exchange of pieces) can not be accurately evaluated \cite{shannon_programmingcomputer_1950}. In order to accurately evaluate a position, the computer performs an extensive search over all possible continuations of the position to a significant extent and selects the one that maximizes the minimum gain \cite{shannon_programmingcomputer_1950}. Such brute force number-crunching is in contrast with a human way of playing chess with creativity and intuition \cite{kasparov}. Here, instead, the coordinate is optimized by making the $F_C$ higher. The numerical parameters of $E(p)$ (e.g., the queen's material value) were (iteratively) randomly modified and kept if the modification resulted in a higher $F_C$ (for details see Appendix). The cFEPs for the optimized coordinate $\tilde{E}(p)$ are marginally higher (by 0.3 around x=0, Fig. 2b) than that for the sub-optimal case (Fig. 2a). $\alpha \sim 0.5$ indicates that the dynamics is diffusive and Markovian and that the optimized reaction coordinate completely specifies the dynamics, i.e., is the optimal reaction coordinate.

The chess game and the protein folding \cite{krivov_is_2010}, both illustrate
 useful generic property of the cFEPs: the higher is the profile the more diffusive is the dynamics, i.e., the dynamics is not sub-diffusive  \textit{per se}. Consider two reasonably good reaction coordinates which differ locally but give similar large-scale description of dynamics.  At sufficiently large times scale ($\Delta t_2$), when memory effects due to sub-optimal projection \cite{zwanzig_nonequilibrium_2001} can be neglected and dynamics is Markovian, the coordinates have similar cFEP. The sub-diffusion exponent $\alpha$ can be estimated from the distance between the profiles computed with the large time step ($\Delta t_2$) and the original (small) time step ($\Delta t_1$):
$$\alpha(x)=1+\frac{\ln Z_C(x,\Delta t_1)-\ln Z_C(x,\Delta t_2)}{\ln \Delta t_1 -\ln \Delta t_2}.$$ The higher is the cFEP (the smaller $Z_C$) at the original time scale of $\Delta t_1$, the smaller is the distance between the profiles, the larger is $\alpha$  and more diffusive is the projected dynamics.

\section*{The equilibrium FEP.}
The $Z_C^+(x)$ and $Z_C^-(x)$ (Fig. 2b) computed from the non-equilibrium trajectories do not represent the underlying (equilibrium) FEP. The latter, however, can be computed from them (Fig. 2c) (see Appendix).  In order to emphasize the robustness of the results, they are recomputed with the (complementary) Markov network formalism (see Appendix), which describes the dynamics by a network of transitions between different states \cite{noe_transition_2008}. Though the profile computed with the Markov network shows some noise, the results obtained with both frameworks are in very good agreement. 

The chess game (as a whole) is described in a simple, while accurate way as diffusion (with $D(x)=1$) on the profile. A game represents a particular realization of the stochastic diffusion process. It starts at $x=0$ and continues until either the right (white wins) or the left (black wins) end of the profile has been reached. The relatively flat region of the profile ($|x|<2.5$) suggests that, initially, a trajectory (game) may switch many times between positive and negative parts. It describes the constructive, search phase of the game, where the opponents are trying to get an advantage. After the barriers ($|x|>2.5$) the profile is much steeper, making a return to the opposite part very unlikely, meaning that the barriers are the rate limiting step in the game dynamics. As soon as a decisive advantage has been gained (a barrier has been overcame) the game strategy becomes much simpler: it is sufficient to exchange the pieces, while keeping the advantage. The probability of overcoming a barrier can be roughly estimated as half of the probability of being at the top of the barrier $p_i\sim 0.5\exp(-F_i/kT)$, where $F_i$ is the barrier height \cite{landau_statistical_1980}. The winning probability (of white) $P=p_w/(p_w+p_b)=e^{-\Delta F/kT}/(1+e^{-\Delta F/kT})\sim 0.6$ is in agreement with the number computed directly from the games of $0.59$; $\Delta F/kT=F_w/kT-F_b/kT \sim -0.4$ (Fig. \ref{joint}c).

\begin{figure*}[htbp]
\centering 
\resizebox*{2.0\columnwidth}{!}{\includegraphics*[]{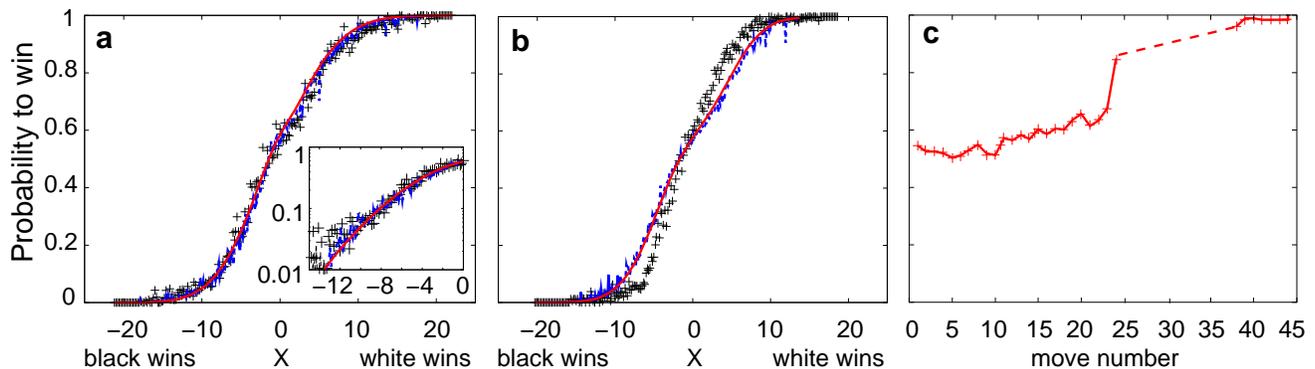}}

\caption{ (color online) \textbf{The probability to win.} \textbf{a)}  The probability (for white) to win the game starting from any position computed along the optimal coordinate $\tilde{E}(p)$: with diffusion on the cFEP (solid red) and with the Markov network framework (dashed blue), and directly from the played games (crosses). The inset shows the plot in logarithmic scale. \textbf{b)} that along the suboptimal reaction coordinate $E(p)$. \textbf{c)} The analysis of the game between Garry Kasparov and Veselin Topalov: probability to win vs the move number.}
\label{win}
\end{figure*}

\section*{The probability to win and game analysis.}
A more stringent test is to compare the winning probability $P(x)$ for any position $x$, which is the probability to reach the right end of the profile, before reaching the left end. It is analogous to the folding probability or commitor used in protein folding studies \cite{onsager_initial_1938, du_transition_1998}.
Fig. \ref{win}a shows $P(x)$ calculated directly from the trajectory, from the diffusive dynamics on the cFEP and with the Markov network framework (for details see Appendix). They are in a very good agreement. $P(x)$ computed with the sub-optimal coordinate (Fig. 2a) notably disagree (Fig. 3b) because dynamics along this coordinate is neither diffusive nor Markovian. Naive approach to estimate the winning probability by collecting statistics for every position is impractical due to the sheer size of the configuration space (estimated as $10^{43}$  \cite{shannon_programmingcomputer_1950} ). Markovian coarse-graining of the space is essential, and is obtained here by the projection onto the optimal coordinate. 
Once constructed, the optimal reaction coordinate can be used to analyze a chess game in simple terms by showing the evolution of the winning probability. As an example, game between Garry Kasparov and Veselin Topalov played in Hoogovens in 1999 is considered (Fig. 3c). Initially, black equalizes the chances to win the game (at 10-th move). Then, white increases the chances to win up to $P=0.7$ (at 23-rd move). The part of the game between 24-th and 38-th moves (dashed line) which starts with the rook sacrifice can not be analyzed due to the shortcomings of the evaluation function. More sophisticated variants of the evaluation function should make possible the analysis of the entire game. 

The optimization procedure effectively tries to decrease the size of each step, so that 
the chess dynamics along the optimal reaction coordinate consists of small incremental changes (e.g., moves 1-23 on Fig. 3c). However, the conclusion that 
a "brilliant" move, which alone can change the course of a game is impossible, is wrong.
Since the game is described as a random process, there is non-zero Gaussian probability to have a large move along the reaction coordinate $p(\Delta x)\sim e^{-\Delta x^2/2}$ (the diffusion coefficient $D=1$). 

\section*{Concluding discussion.}
An optimization (learning) principle to simultaneously optimize the reaction coordinate and the playing strategy can be suggested. $Z_C^-(x)$ (Fig. 2b, blue) which counts the "retrograde" moves, is an indicator of the quality of play. The quality consistently improves with increasing $x$ since the complexity of the game decreases. When moves are chosen according to the "perfect" winning strategy $Z_C^-(x)=0$. Hence, $F_C(x)$ attains the upper bound ($Z_C^-(x)=0$) for the best playing strategy and the optimal coordinate describing it. The principle can be used to measure and optimize performance of stochastic algorithms, for example, to improve global optimization heuristics \cite{kirkpatrick_optimization_1983}. 

The presented analysis can be improved in the following ways: the development of more sophisticated variants of the evaluation function to treat the dynamic phases of the game; the construction of reaction coordinates tailored to different types of positions, or perhaps  reaction coordinate which is iteratively updated during the game; two dimensional free energy landscape to allow analysis of games ending in a draw. 

Assuming that the chess game is a stochastic process the optimal coordinate that provides accurate description of the process has been constructed. The cut free energy profile is a function of the reaction coordinate time series alone. Detailed specification of the dynamics (even the rules of the game) is not necessary to perform the optimization. It allows the approach to analyze phenomena which numerous characteristics can be monitored, while construction of the complete dynamical model is impractical. The analysis of the chess game shows that the approach can quantitatively describe the dynamics of processes where human decision-making plays a central role, e.g., financial and social dynamics.

An interesting application is the construction of disease biomarkers \cite{poste_bring_2011} 
, specifically for such difficult cases as cancer \cite{sawyers_cancer_2008}, ageing \cite{vaupel_biodemography_2010} or psychiatric disorders \cite{singh_biomarkers_2009}. An optimal biomarker and the associated free energy landscape that give an accurate description of the disease dynamics can be constructed out of panel of monitored parameters e.g., metabolomic, proteomic and genomic data.  Diffusive dynamics would indicate that the optimal biomarker gives complete (Markovian) description, and sub-diffusive that some essential information is missing. Unlike conventional approaches, the proposed approach explicitly treats the dynamical character of the process.

\begin{acknowledgments}
This work was supported by an RCUK fellowship.
\end{acknowledgments}

\section*{Appendix}

\textbf{The cut free energy profiles.}
Given reaction coordinate ($R(\vec{X})$) and an ensemble of trajectories $X_i(j\Delta t)$ sampled with interval $\Delta t$ the ensemble of reaction coordinate trajectories is defined as $x_i(j\Delta t)=R(\vec{X}_i(j\Delta t))$. The partition function of the conventional (histogram-based) free energy profile is estimated as 
\begin{eqnarray}
Z_H(x)=N_x/\Delta x,
\end{eqnarray}
where $N_x$ is the number of trajectory points in bin $x$ and $\Delta x$ is the size of the bin. 
The partition function of the cut based free energy profile \cite{krivov_diffusive_2008} at point $x$ is estimated as half the number of transitions through that point, i.e., 
\begin{eqnarray}
Z_C(x)=1/2\sum_{i,j} \Theta\{(x_i(j \Delta t)-x)(x-x_i(j \Delta t+\Delta t)\},
\end{eqnarray}
where $\Theta\{x\}$ is the Heaviside step function. 

For Gaussian distribution of steps $P(\Delta x)\sim \exp(-\Delta x^2/(4D\Delta t))$ and assuming that $F_H(x)$ is approximately constant on the distance of the mean absolute displacement $\langle |\Delta x(\Delta t) |\rangle$, one obtains \cite{krivov_diffusive_2008}
\begin{eqnarray}
Z_C(x)=Z_H(x)\sqrt{D(x)\Delta t/\pi};
\label{eq01}
\end{eqnarray}
For non-Gaussian distribution of steps, observed here, assuming $\langle \Delta x^2\rangle=2D\Delta t$, Eq. \ref{eq01} is corrected by factor $\nu$
\begin{eqnarray}
Z_C(x)=\nu Z_H(x)\sqrt{D(x)\Delta t/\pi}\\ \nu=\langle |\Delta x|\rangle/\sqrt{2\langle \Delta x^2\rangle/\pi} 
\label{eq02}
\end{eqnarray}
The correction factor $\nu$ is assumed to be coordinate independent and is computed from the entire trajectory. The distribution of steps is expected to converge to Gaussian for more sophisticated evaluation functions, where each move changes a significant number of terms of the function.

When dynamics is not equilibrium the net trajectory flux is not zero and the detailed balance is not satisfied. In this case positive and negative cut profiles that measure the flow in the corresponding direction are introduced
\begin{eqnarray}
Z_C^+(x)=\sum_{i,j} \Theta(x-x_i(j \Delta t))\Theta(x_i(j \Delta t+\Delta t)-x) \\
Z_C^-(x)=\sum_{i,j} \Theta(x_i(j \Delta t)-x)\Theta(x-x_i(j \Delta t+\Delta t)).
\end{eqnarray}

\textbf{Diffusion with constant flux \textit{J} and gradient \textit{F(x)=ax}.}
The flux for the steady state solution $P_{st}(x)$ of the Smoluchowski equation is
$$J=-D(x)e^{-\beta F(x)}\partial/\partial x (P_{st}e^{\beta F(x)}),$$
where $\beta=1/kT$ and $D(x)=D$.
Steady state distribution is found as $$P_{st}(x)=Z_H(x)=\frac{J}{D\beta a}-Ce^{-a\beta x},$$ where C is a constant defined by a boundary condition.
Distribution of displacements $\Delta x$ during time step of $\Delta t$ is $$p(\Delta x,\Delta t)=\frac{1}{\sqrt{4\pi D \Delta t}}\exp[-\frac{(\Delta x +Da\beta \Delta t)^2}{4D\Delta t}].$$
$Z_C^+$, the number of transition from $y<0$ to $y>0$ equals to
$$Z_C^+=\int_{-\infty}^0 dy Z_H(y) \int_{-y}^{\infty}p(x,\Delta t)dx.$$
$Z_C^-$, the number of transition from $y>0$ to $y<0$ equals to
$$Z_C^-=\int_0^{\infty} dy Z_H(y) \int^{-y}_{-\infty}p(x,\Delta t)dx.$$
Integrating over y one obtains
$$Z_C^+=\int_0^{\infty} dx p(x,\Delta t) (\frac{Jx}{D\beta a} +C\frac{1-e^{\beta a x}}{\beta a})$$
$$Z_C^-=\int_{-\infty}^0 dx p(x,\Delta t) (-\frac{Jx}{D\beta a} -C\frac{1-e^{\beta a x}}{\beta a}).$$
The net flow equals to
$$Z_C^+-Z_C^-=\int_{-\infty}^{\infty} dx p(x,\Delta t) \frac{Jx}{D\beta a}=J\Delta t$$
$Z_C=(Z_C^++Z_C^-)/2$ is found by expanding the exponents
$$Z_C=Z_H\sqrt{D\Delta t/\pi}(1+O(D\Delta t\beta^2a^2))$$

\textbf{The equilibrium FEP.}
$F(x)$ can be found as
$$F(x)/kT=-\ln P_{st}(x) -\int^x \frac{J(x)dx}{D(x)P_{st}(x)}.$$ Using $P_{st}(x)=Z_H(x)$, $J(x)\Delta t=Z_C^+(x)-Z_C^-(x)$ and $D(x)=(Z_C/Z_H)^2\nu^{-2}\pi/\Delta t$, one obtains 
\begin{eqnarray}
F(x)/kT=-\ln Z_H(x) -\int^x\frac{\nu^2 (Z_C^+(x)-Z_C^-(x))Z_H(x) dx}{\pi Z_C^2(x)}
\label{eq1}
\end{eqnarray}
\textbf{The natural coordinate.} A reaction coordinate (x) with variable diffusion coefficient can be transformed to the natural coordinate (y) with the diffusion coefficient equals to unity by numerically integrating \cite{krivov_diffusive_2008}
\begin{eqnarray}
dy=\nu^{-1} \sqrt{\Delta t/\pi}Z_h(x)/Z_c(x)dx. 
\label{eq3}
\end{eqnarray}
In this case the diffusive dynamics is completely specified by the free energy profile only. 

\textbf{The sub-diffusion exponent.} In case of sub-diffusive dynamics, the mean absolute displacement scales with time as $\langle |\Delta x(\Delta t) |\rangle  \sim \Delta t^{\alpha}$, where $\alpha<0.5$. The coordinate dependent exponent $\alpha$ is determined by comparing $Z_C(x)$ at two different sampling intervals \cite{krivov_is_2010}
\begin{eqnarray}
\alpha(x)=1+\frac{\ln Z_C(x,\Delta t_1)-\ln Z_C(x,\Delta t_2)}{\ln \Delta t_1 -\ln \Delta t_2}. \label{alpha_eq}
\end{eqnarray}

\textbf{Reaction coordinate optimization.} The optimal reaction coordinate is defined as the one that has cut based free energy profile $F_C(x)$ highest for every value ($x$) of the coordinate \cite{krivov_one-dimensional_2006,krivov_diffusive_2008, krivov_is_2010}. It generalizes the definition of the transition state as the minimum cut
to any position on the reaction coordinate \cite{krivov_one-dimensional_2006}.
Projection on a ''bad'' reaction coordinate results in smaller barriers and faster kinetics due to overlapping of different parts of the configuration space. The optimal coordinate exhibits the slowest kinetics \cite{krivov_diffusive_2008}. The dynamics projected on this coordinate is closest to diffusive \cite{krivov_is_2010}.

The coordinate as a whole is optimized by numerically maximizing $\int Z_C^{-1}(x)Z_H(x)dx$. For a flexible form of the reaction coordinate where $Z_C(x)$ for all $x$ can be considered as independent the functional attains the maximum when all $Z_C(x)$ are minimal (all $F_C(x)$ are maximal). For a less flexible reaction coordinate the optimum may be a compromise solution where different parts of the coordinate are optimized to a different degree.  Another optimization functional $\max \int Z_C^{-2}(x) Z_H(x) dx \sim \int e^{F(x)/kT}/D(x)dx$, which assigns more weight to the higher part of the profile, gives very similar results. For the case of over-damped Langevin dynamics $$\int e^{F(p)/kT}/D(p)dp$$ attains the maximum when $p$ is the committor function ($p^{comm}(\vec{X})$) or the folding probability, an alternative definition of the optimal reaction coordinate \cite{we_metastability_2004}. 

The reaction coordinate is a weighted sum of different components of the evaluation function $E(p)=\sum a_i f_i(p)$, where $a_i$ are the weights, $f_i(p)$ are the components of the evaluation function and $p$ denotes the position.  Starting with the initial set of the weights $a_i^0$ (that of the GNUCHESS program) the coordinate is iteratively improved by randomly changing the weights $a_i$ and accepting the change if the value of the functional is increased.

The cFEP is invariant with respect to local changes of scale of the evaluation function (gauge invariant), it depends only on the relative order of the points. This makes the construction of the optimal coordinate much easier, since a putative reaction coordinate $E(p)$ should only reproduce the relative order, not the absolute value of the goodness of the positions. 

\textbf{The Markov network framework.}
The Markov network describing the dynamics is constructed by partitioning the reaction coordinate into bins of size 0.005 and computing the transition probabilities as $p_{ij}=n_{ij}/\sum_i n_{ij}$, where $n_{ij}$ is the number of transitions from bin $j$ to bin $i$. The reaction coordinate was converted to the natural coordinate to ensure uniform partition. 
\textit{The equilibrium FEP} was constructed by computing the equilibrium network $n_{ij}^{eq}$ 
\begin{eqnarray}
n_{ij}^{eq}=p_{ij}p_j^{eq},
\end{eqnarray}
where $p_j^{eq}$ are the equilibrium populations 
\begin{eqnarray}
p_i^{eq}=\sum_j p_{ij}^{\prime} p_j^{eq},
\end{eqnarray}
 and the sub-network $p_{ij}^{\prime}$ is the strongly connected component of the network, i.e., there is a path in each direction between any two nodes of the sub-network. In particular, the terminal (checkmate) nodes are discarded. 
$Z_C$ and $Z_H$  are computed as 
\begin{eqnarray}
Z_C(x)&=&1/2\sum_{ij} n_{ij}^{eq}\Theta\{(x(i)-x)(x-x(j))\}\\
Z_H(x)&=&\sum_{ij,x(j)=x} n_{ij}^{eq},
\end{eqnarray}
where $x(i)$ and $x(j)$ are the positions of the clusters $i$ and $j$, respectively.

\textbf{The probability to win} (the committor  or the folding probability in the analysis of protein folding dynamics) measures probability to reach point $x=b$ before reaching point $x=a$. For the diffusion on a one-dimensional free energy profile it can be estimated \cite{rhee_one-dimensional_2005} as $$p^{comm}(x)=\int_a^x e^{F(x)/kT}/D(x)dx/\int_a^b e^{F(x)/kT}/D(x)dx,$$ which can be expressed as
\begin{eqnarray}
p^{comm}(x)=\frac{\int_a^x Z_h(x) Z_c^{-2}(x)dx}{\int_a^b Z_h(x) Z_c^{-2}(x)dx}
\label{eq2}
\end{eqnarray}
It can be estimated from the Markov network as
\begin{eqnarray}
p^{comm}_i=\sum_jp_{ji}p^{comm}_j
\label{eq4}
\end{eqnarray}
with $p^{comm}_{1-0}=1$ and $p^{comm}_{0-1}=0$, where 1-0 and 0-1 are the nodes corresponding to the states where white or black has won, respectively. 
It can be estimated directly from the trajectories as
$p^{comm}_i=n^w_i/(n^w_i+n^b_i)$, where $n^w_i$ and $n^b_i$
are the number of times trajectory visiting bin $i$ ends in white's or black's victory, respectively \cite{rao_estimation_2005}.

\textbf{Preprocessing of the chess game trajectories.}
The ensemble of the chess game trajectories consists of 10000 games played by the GNUCHESS program against itself with default parameters. The positions during highly dynamic phases of the game (e.g., during exchange of pieces) can not be accurately evaluated with the employed evaluation function \cite{shannon_programmingcomputer_1950}. To alleviate these problems the trajectories were processed as follows. Initially, every continuous  sequence of exchanges of pieces is cut off from the game trajectories. The trajectories were projected to the evaluation function coordinate by computing the evaluation function for every second position with white turn to move (every fourth ply;  ply is a single move either by white or black). The trajectories then were transformed (normalized) to the natural coordinate. To avoid highly dynamic phases of the game other than the continuing sequence of exchanges of pieces, the steps with change in the value of the reaction coordinate larger than the threshold value of 5 (along the natural coordinate) were discarded. The first two moves of every game were discarded, to remove the very early development phase of the game and keep the game dynamics homogeneous. More sophisticated variants of the evaluation function should make these preprocessing steps unnecessary.

%
\end{document}